\documentclass{ws-ijmpcs}
\usepackage{epsf}

\begin{document}

\markboth{Bisnovatyi-Kogan G.S., Klepnev A.S., Giovannelli F.}
{Estimation of alpha-viscosity coefficient from observations of nonstationary disk accretion}

\catchline{}{}{}{}{}
%
%%%%%%%%%%%%%%%%%%%%%%%%%%%%%%%%%%%%%%%%%%%%%%%%%%%%%%%%%%%%%%%%%%%%

\title{ESTIMATION OF ALPHA-VISCOSITY COEFFICIENT FROM OBSERVATIONS OF NONSTATIONARY DISK ACCRETION}

\author{BISNOVATYI-KOGAN G.S., KLEPNEV A.S.}

\address{Space Research Institute, Profsoyuznaya 84/32,
Moscow 117810, Russia; and \\ National Research Nuclear University MEPhI,
Kashirskoe Shosse 31,\\ Moscow 115409, Russia
}

\author{GIOVANNELLI F.}

\address{INAF - Istituto di Astrofisica e Planetologia Spaziali -
Area di Ricerca di Tor Vergata - \\Via del Fosso del Cavaliere 100, I
00133 Roma, Italy}

\maketitle

\begin{history}
\received{Day Month Year}
\revised{Day Month Year}
\end{history}

\begin{abstract}
The optical behaviour of the Be star in the high mass X-ray transient A0535+26/HDE245770 shows that at
the periastron typically there is an enhancement in the luminosity of order 0.05 to few tenths mag, and the X-
ray outburst happens about 8 days after the periastron. We construct a quantitative model of this event,
basing on a nonstationary accretion disk behavior, connected with a high ellipticity of the orbital motion. We
explain the observed time delay between the peaks of the optical and X-ray outbursts in this system by the
time of radial motion of a matter in the accretion disk, after increase of the mass flux in the vicinity of a
periastral point in the binary. This time is determined by the turbulent viscosity, with the parameter $
\alpha=0.1-0.3$, estimated from the comparison of the model with observational data

\end{abstract}

%\ccode{PACS numbers: 11.25.Hf, 123.1K}

\section{Introduction}	

The X-ray source A0535+26 was discovered by Ariel V satellite on 14 April, 1975. The X-ray source was in outburst with the intensity of $\approx 2$ Crab, and showed  pulsations \cite{ariel} with a period $\sim 104$ s.
The Be star HDE 245770 was discovered \cite{2} as the optical counterpart of A0535+26, and was classified \cite{16} as O9.7IIIe star.
A review of this system can be found in Ref.8.
Briefly, the properties of this systems, placed at distance \cite{16} of $1.8 \pm 0.6$ kpc, can be summarized as follows: hard X-ray transient, long period X-ray pulsar -- the secondary star -- orbiting around the primary O9.7IIIe star. The masses are \cite{{21},{33},{22}} of $\sim 1.5 \pm 0.3$ M$_\odot$, and 15 M$_\odot$ \cite{16} for the secondary and primary stars, respectively. The eccentricity is e = 0.497 \cite{gl92}. Usually the primary star does not fill its Roche lobe \cite{dlgd}.
We assume here the orbital period \cite{29a} obtained from X-ray data: P$_{\rm orb} = 111.0 \pm 0.4$ days.

 The 111-day orbital period is in agreement, inside the error bars,  with many other determinations reported in the literature.
It was stressed \cite{fg} that
 the epoch of the periastron passage of the neutron star around the Be star is always before X-ray outbursts of $\sim 8$ days. Here we construct the model explaining this feature, which permits to estimate the viscosity $\alpha$ - parameter, as $\alpha= 0.1\,-\, 0.3$.
 The detailed description of the observational data and of the model is given in Ref. 10

\section{Observational evidences of the time delay}

Here we present  observational data that demonstrate the 8 days delay between the relative enhancement of the optical luminosity of HDE 245770, occurring at the periastron passage, and the consequent X-ray outburst of the X-ray pulsar A0535+26, see more in Ref. 10.
Fig.1 (left) shows the optical relative maximum luminosity in U, B, and V bands of HDE 245770 occurred on December 5, 1981, and the date (December 13, 1981) when the subsequent X-ray outburst occurred.

\begin{figure}[pb]
\centerline{\psfig{file=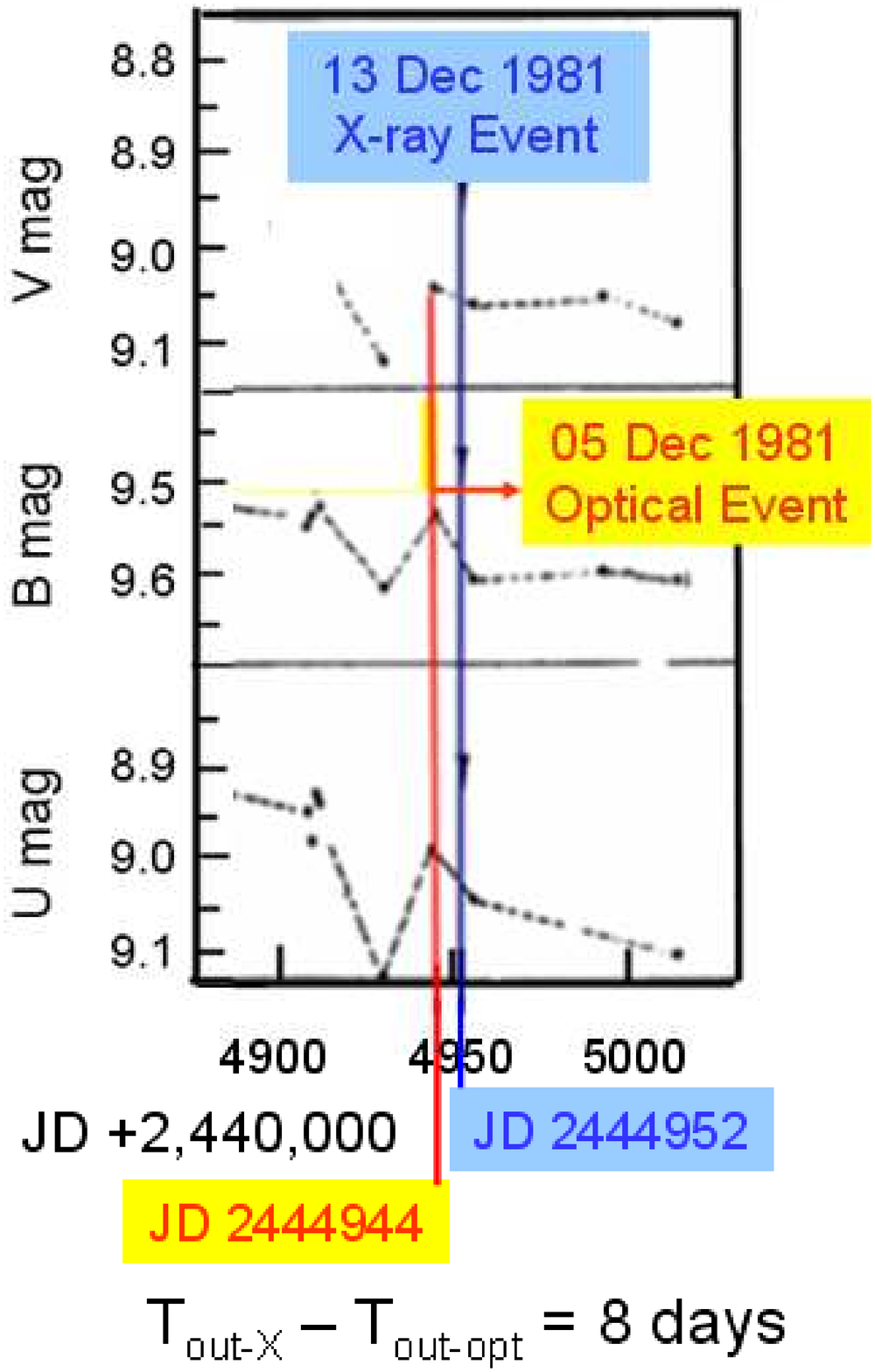,width=6cm}
            \psfig{file=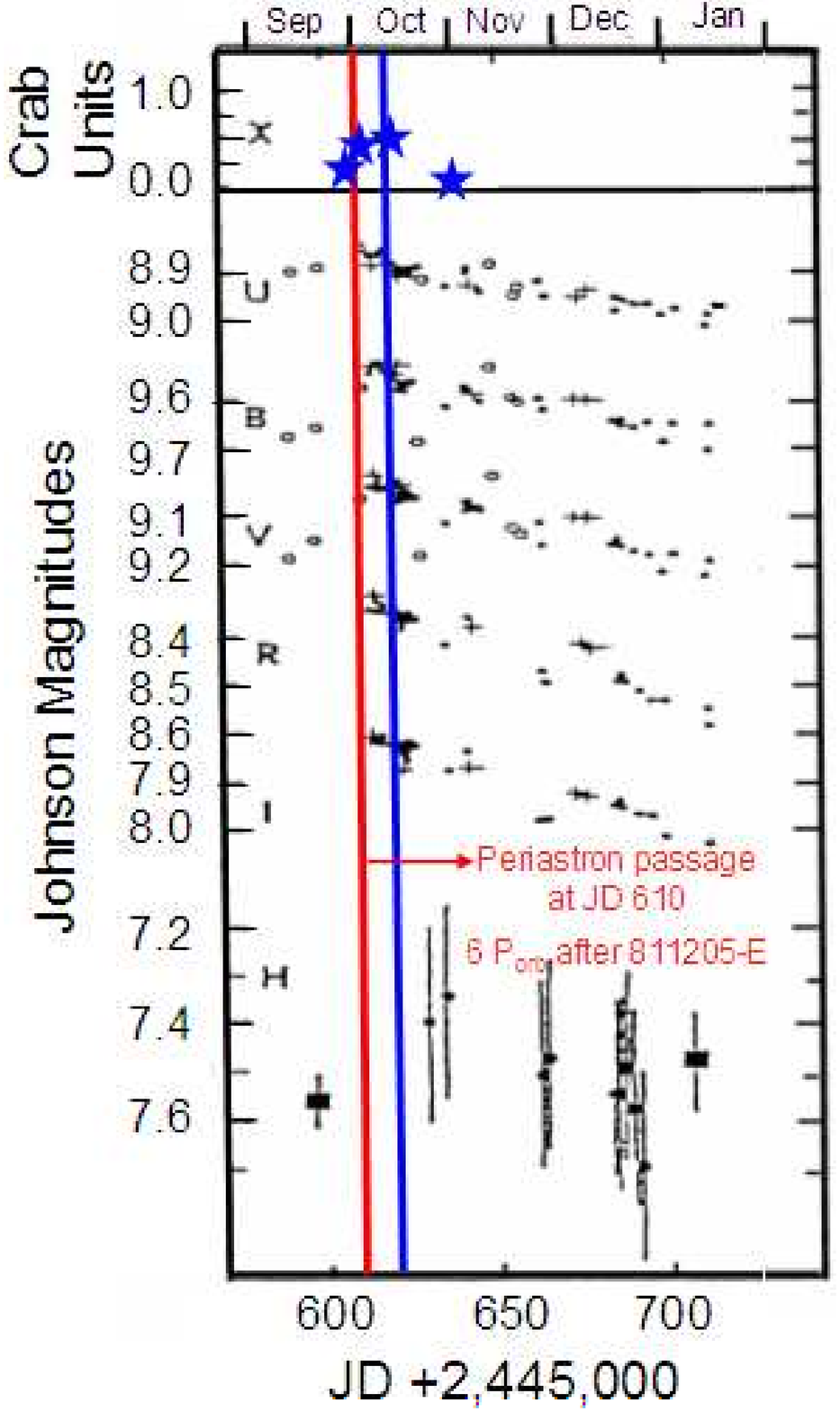,width=6cm}}
\caption{Left panel. Optical outburst on December 5, 1981 
 occurred
8 days before the short and sharp X-ray outburst of December 13,
1981.
Right panel. Optical outburst in U,B,V,R,I, and H bands at the 6th periastron passage after the previous one.
Optical maximum coincides with the periastron passage, X-ray data  are marked with  stars; the nearby line marks the maximum of the X-ray outburst.
}  \label{f3}
\end{figure}
%%%%%%%%%%%%%%%%%%%%%%%%%%%%%%%%%%%%%%%%%%%%%%
After six orbital periods the optical flare in U,B,V,R,I, and H bands was found  just $\sim 8$ days before the maximum of the X-ray outburst occurred in October 1--18, 1983 \cite{fg}. Fig.1 (right) shows these events. The left line marks the periastron passage, coinciding with the optical maximum, and the nearby line marks the maximum of the X-ray outburst. The stars represent observational points \cite{fg}.

\section{Properties of the X-ray transient A0535+26/HDE245770}

The observed orbital period $P$ and eccentricity $e$ are
$P \simeq 111\,\,{\rm days} \hfill, e = 0.47$.
From the Kepler law we have a relation
$P = 2\pi {a^{3/2}}\sqrt {\frac{1}{{G(M + m)}}}$,
  where $M=14\, M_\odot$ and $m=1.4\, M_\odot$ are the masses of the optical and neutron stars, respectively. The large semiaxis of the orbit $a$, and minimal separation of the stars at the periastron $r_{\min}$ from equation (1) are determined as 
\begin{equation}
 a = \sqrt[3]{{\frac{M}{{{M_{\odot}}}}}}{(\frac{P}{{{P_{\rm Earth}}}})^{2/3}}\sqrt[3]{{1 + \frac{m}{M}}}\approx 1.8 \cdot 10^{13}\, {\rm cm},
%  \label{eq2}
% \end{equation}
%
%\begin{equation}
  \quad {r_{\min }} = a(1 - e)\approx 9.6 \cdot 10^{12}\, {\rm cm}.
  \label{eq3}
 \end{equation}
where $P_{\rm Earth}= 1$ year.
 In quasistatic approximation,  a radius of the Roche lobe around the neutron star $r_{roche}$, at periastron, is found in Ref.6.

\section{Model of the time delay}

In the vicinity of the periastron the mass flux $\dot M$ rapidly increases.
 The outer parts of the accretion disk become hotter, increasing the optical luminosity. Due to turbulent viscosity, the wave of the large mass flux is propagating to the central object. When this wave reaches the vicinity of the neutron star the X-ray luminosity increases due to appearance of a hot accretion disk region, and due to luminosity of magnetic poles, where the gas flows due to magnetic field channeling \cite{{bkf69},{bk10}}.
We identify the time delay $\tau$ between the optical and X-ray flashes, with the time during which the wave of a high mass accretion flux, starting from  $r_{\rm out}$, reaches the central compact star radius $r_{\rm in}$. The speed of this wave is taken to be equal to the local radial speed $v_r$ of the matter, corresponding to the local mass flux.
 We consider a geometrically thin, optically thick accretion disk without advection, and
suggest, that at each radius, with $\dot M(r,t)$, the accretion disk parameters are the same as in the stationary accretion disk with the same $\dot M$ over the whole disk.
 The system of equations describing such disk,
 consist of \cite{{ss73},{bk10}}
 the mass conservation equation
$\dot {M}=4\pi rh\rho v_r$;
the angular momentum equation 
$\frac{\dot {M}}{4\pi }\frac{dl}{dr}+\frac{d}{dr}(r^2ht_{r\varphi })=0$,
which after integration it reduces to 
$r^2ht_{r\varphi } =-\frac{\dot {M}}{4\pi }(l-l_{in} )$, where
$\ell=\Omega r^2$  is a specific angular momentum, and $ t_{r\phi}=-\alpha P$ is the ($r,\phi$) component of the viscosity tension tensor \cite{ss73}. 
$\Omega$ is the Kepler angular velocity $\Omega^2=GM/r^3$. The value of the $l_{in}$ is determined by the angular momentum at the inner boundary of the accretion disk, with zero derivative of the angular velocity. For the accretion into a non-magnetized, slowly rotating  neutron star $r_{\rm in}$ is close to its radius $R_{\rm ns}$ \cite{bk10}.  In the case of a strongly magnetized star, where the Alfv\'{e}n radius
 $r_A\gg R_{\rm ns}$ a definition of the inner radius and inner angular momentum are less clear, because the matter flows along the field lines to magnetic poles from the Alfv\'{e}n surface. In our problem the time delay is determined by a slow radial motion in the outer parts of the accretion disk, where  $l\gg l_{in} $ and therefore the choice of the value of $l_{\rm in}$ is not important. 
The local equation of the energy conservation is written as
$Q^+=Q^-\,\, {\rm (erg/cm^2/s)}$,
where $Q^+=ht_{r\varphi } r\frac{d\Omega }{dr}$ is the energy production rate by a viscous dissipation, related to the unit of the disk surface,
 and $Q^-=\frac{2acT^4}{3\tau _0}$ is a radiative flux from the optically thick disk, through the unit of the disk surface.
 Here $T$ is the temperature, $a$ is the constant of the radiation density, and $\tau_{0}=\kappa \rho h$ is the Thompson optical depth, given by $\tau_{0}=0.4 \rho h$ for a hydrogen composition.
The pressure $P_{\rm tot}=P_{\rm gas}+P_{\rm rad}=\rho{\cal R}T\,+\,{a T^4 \over 3}$ is determined by a mixture of matter $P_{\rm gas}$ and radiation $P_{\rm rad}$, 
 ${\cal R}$ is the gas constant.
The system of equations is reduced to a single algebraic non-linear equation for the sound speed $c_s$ \cite{art96}
Solving this equation we find $c_s$. Radial velocity of the matter in the disk is written as
$  {v_r} = \frac{{\alpha c_s^2}}{{r\Omega f}}$.
The  parameters in our problem are the viscosity parameter $\alpha$, and the initial mass flux function $\dot M(r)$.
The mass of the neutron star is supposed to be known from observations.
The initial distribution of the matter flux in the disk is defined as
$\dot{M}(r)=\frac{r_{\sigma} ^{2} }{(r-r_{0} )^{2} +r_{\sigma} ^{2} }{\dot M_0}$.
Here $r(t)$ define the Lagrangian radius of the disk matter, $r_0=1.1\cdot 10^{11}$\,cm is a point of the maximum mass flux over the disk at the initial time, and $r_{\sigma}=3.2\cdot 10^{10}\,$cm is a parameter, determining the
 characteristic width the high mass flux region. We make calculations for $\dot M_0 = 10^{-7}
 M_\odot$/year,  $\dot M_0 = 3\cdot 10^{-8}
 M_\odot$/year and $\dot M_0 = 10^{-8}
 M_\odot$/year \cite{fg}.
Knowing the local radial velocity $v_r(r,t)$, 
 we  find the time evolution of the initial distribution of
matter, in the time interval $ dt $ by the formula
$ r(t+dt)=r(t)-v(r)dt$.
Knowing how the matter in the disk is moving, we calculate  the variation of the luminosity  with time, in the optical and  X-ray ranges. The luminosity  consists of two components: emission from the disk and emission from hot spots on the magnetic poles.
The integrated optical luminosity of the disk was calculated  in the range 300--700 nm, and for the luminosity in the X-rays band it was done  in the range  2--10 keV.

Hot spots are formed at the magnetic poles of the neutron star
due to infall of the  matter from the disk along the magnetic field lines.
The angular size of the bottom of the column on the neutron star surface is approximately defined \cite{bk10} by  the  expression $\sin ^{2} \theta =\frac{r_{ns} }{r_{A} }$,
where $r_{ns}$ is the radius of the neutron star, and $r_{A}$  is the radius of the Alfv\'{e}nic surface  
at which the magnetic pressure is equal pressure of the matter.
For our system the equatorial magnetic field on the neutron star surface is taken equal to  $B=4\cdot 10^{12}\,$ Gs, $r_{ns}=10\,$km.
Knowing the size of the hot spots, it is possible to estimate its effective temperature
%
%\begin{equation}
$T_{eff}=\left(\frac{\eta \dot{M}c^2}{S\sigma} \right)^{1/4} =2.4\cdot 10^8 \left(\frac{\dot{M}_{7}}{S_{10} } \right)^{1/4} K,
$
%\end{equation}
where $S=\pi r_{ns}^2\left(\frac{r_{ns}}{r_{A}}\right)$ is the surface of the hot spot,
$\eta\approx \frac{r_{g} }{r_{ns} } =0.3$ is the efficiency of a conversion of the kinetic energy of a falling matter into a radiation.

\section{Results}

The results for the light curves of the system in these two bands, together with a bolometric luminosity, are given in Fig.2. 

\begin{figure}[pb]
\vspace{-1cm}
\centerline{\psfig{file=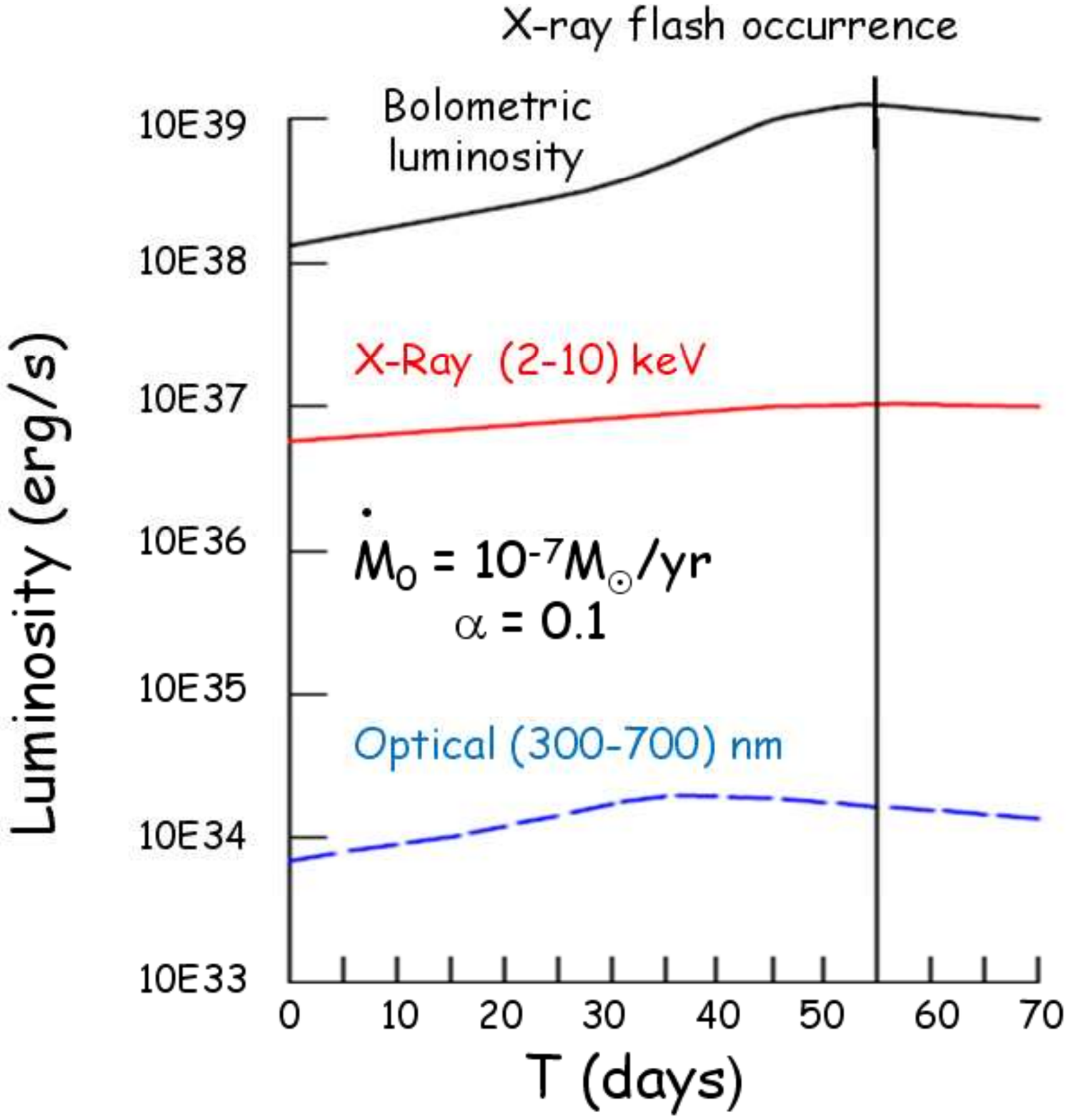,width=4.2cm}
            \psfig{file=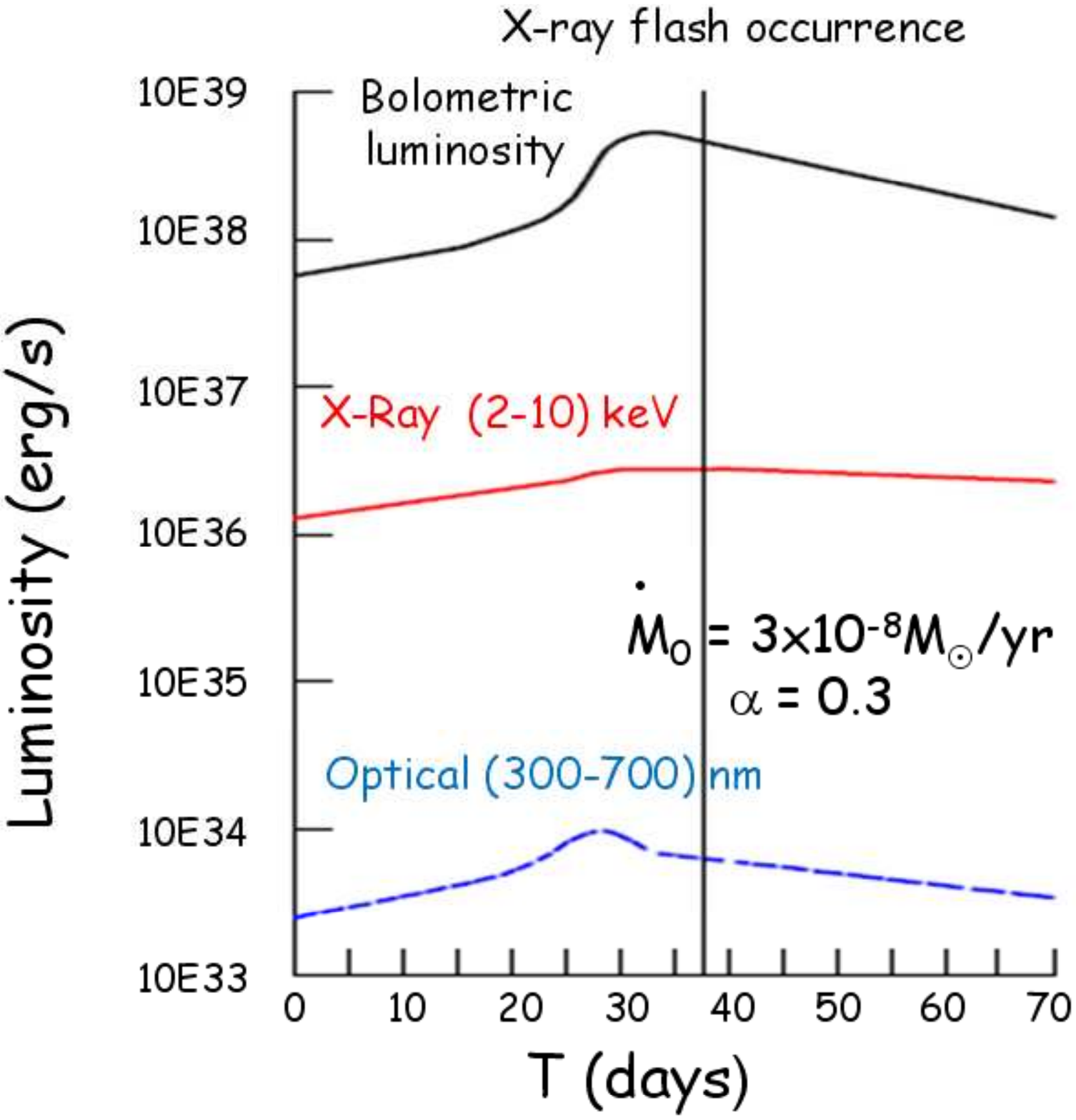,width=4.2cm}
            \psfig{file=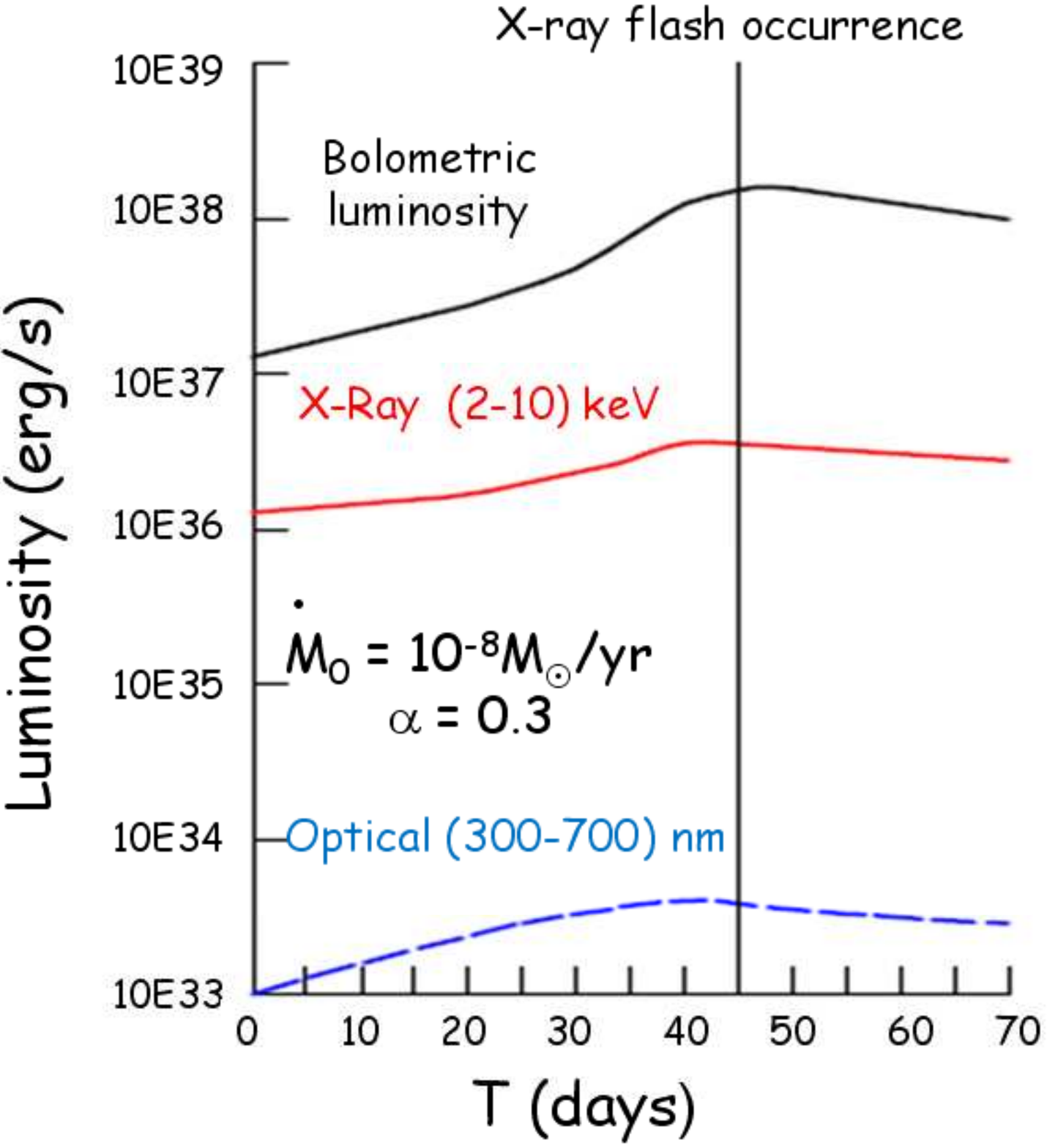,width=4.2cm}
}
%\vspace*{3pt}
\vspace{-1cm}
\caption{The time dependence of the: i) bolometric luminosity of the object (black line), ii) X-rays in the band 2-10 keV (red line),  and iii) optical in the band 300-700 nm (broken blue line). Black vertical line marks the X-ray (2-10 keV) flash occurrence,  about 8 days after the maximum of the optical curve. $\dot M$ is taken from (14), with the following $\dot M_0$.
 Left panel: $\dot M_0=10^{-7}\,M_\odot$/year, $\alpha=0.1$. Middle panel: $\dot M_0=3\cdot 10^{-8}\,M_\odot$/year, $\alpha=0.3$. Right panel: $\dot M_0=10^{-8}\,M_\odot$/year, $\alpha=0.3$.}  \label{f1}
\end{figure}

\begin{figure}[pb]
\vspace{-1cm}
\centerline{\psfig{file=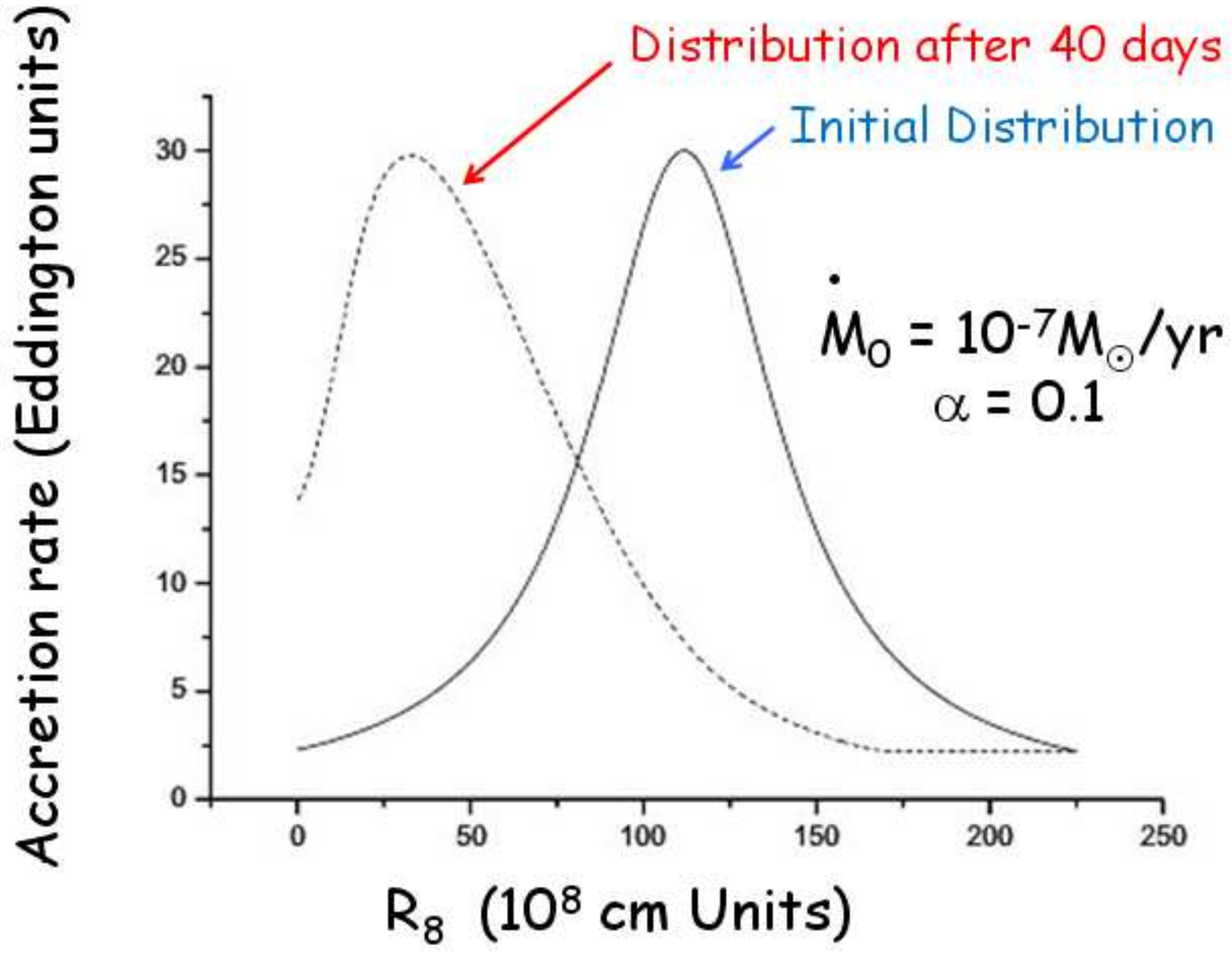,width=4.2cm}
            \psfig{file=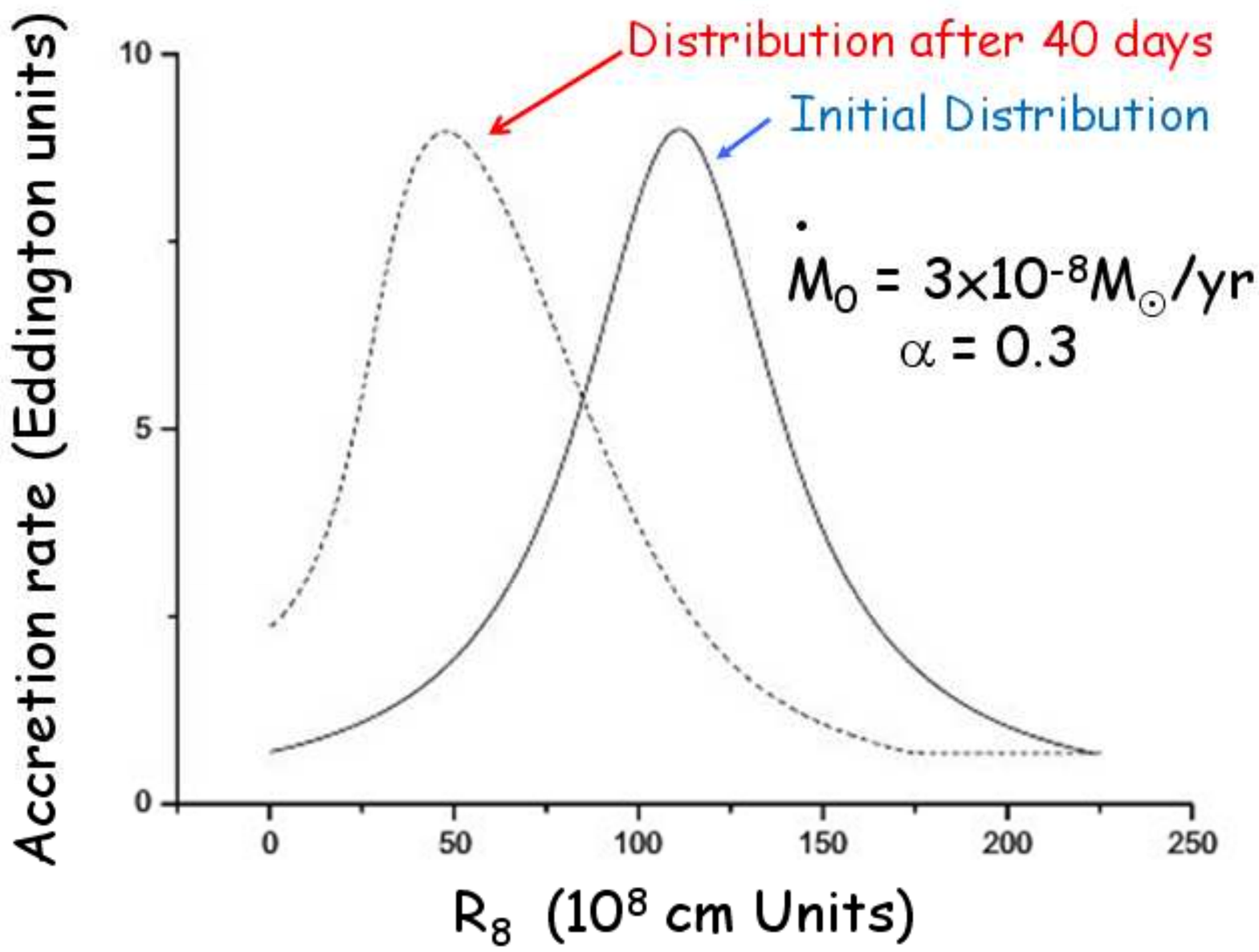,width=4.2cm}
            \psfig{file=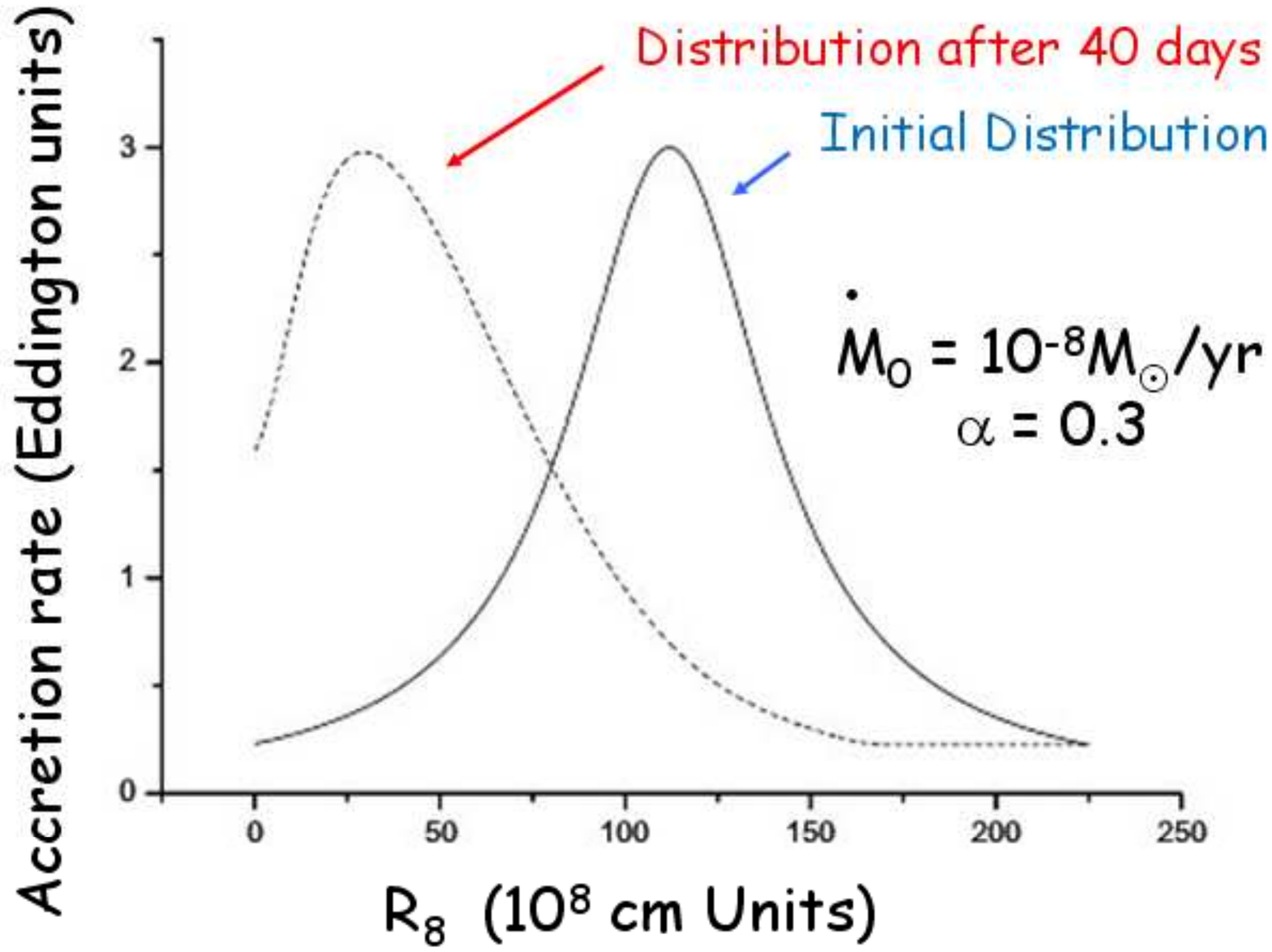,width=4.2cm}
}
\vspace{-1.5cm}
\caption{The radial dependence of the  accretion rate in the disk.
Accretion rate is in the units of the  Eddington  accretion rate, and R$_8$ is in $10^8$ cm units.
 Solid line  shows the initial distribution and, the broken line shows the distribution of accretion rate on the disk after 40 days of evolution.
 $\dot M$ is taken from (14), with the following $\dot M_0$. Left panel: $\dot M_0=10^{-7}\,M_\odot$/year, $\alpha=0.1$. Middle panel:  $\dot M_0=3\cdot 10^{-8}\,M_\odot$/year, $\alpha=0.3$. Right panel:  $\dot M_0=10^{-8}\,M_\odot$/year, $\alpha=0.3$.} \label{f2}
\end{figure}
 The maximum radiation in the X-ray range is reached later than the maximum at optical wavelengths.
The main parameter of the problem that we have been able to vary is the viscosity parameter $\alpha$ for the accretion disk.
We have found that  the time delay between the maxima of the X-ray  and optical emissions  that we need for explaining of the observational data  happens at a value of the  viscosity parameter $\alpha \approx 0.1$ for the flash with $\dot M_0 = 10^{-7} M_\odot$/year, and at a value of  $\alpha \approx 0.3$ for the flashes with $\dot M_0 = 3\cdot 10^{-8} M_\odot$/year, and $\dot M_0 = 10^{-8} M_\odot$/year. 
The disk is formed due to strong increase of the mass flux when the neutron star approaches the periastron of the orbit. The main emission initially comes from outside, relatively cold regions of the disk, and
 the maximum luminosity comes at the optical range.
The contribution of the  disk to the X-ray luminosity at this stage is small. 
When the "accretion wave" in the disk approaches the neutron star, the X ray luminosity increases, and the optical one falls.
The radiation from the hot spots at the poles of the neutron star is added, which  makes a significant contribution to the radiation in the X-rays.
Fig.3 shows the radial dependence of the  accretion rate in the disk for different mass fluxes.

Note that, according to our model, during the falling flow onto the accreting star, the increase of the
luminosity should always start from the low energy part of the spectra, and the maximum in the high energy band follows that in the low energy.
This is contrary to that occurring during the ejection of the hot gas from a star or from the active galactic nucleus (AGN), where the increasing of luminosity starts from the high energy side, and the maximum in the low energy part happens after.

\section{Conclusions}

 The quantitative model of 8 days time delay between optical and X ray maxima in the high mass X ray transient A0535+26 is constructed, basing on a nonstationary accretion disk behavior, due to a high ellipticity of the orbital motion.
For bright outbursts the 8 days delay happens for $\alpha = 0.1$, and for weaker ones $\alpha = 0.3$ is needed.
Our model is valid also for AGNs, where similar type of delays between earlier optical and subsequent X-ray bursts were observed \cite{{n98},{mrm08}}.
Quantitative model of light curves in these outbursts should give  information about properties of a turbulent viscosity in the accretion disks around supermassive black holes.

\section{Acknowledgements}

The work of GSBK and ASK was partially supported by the RFBR grant 11-02-00602, the RAN Program
'Origin, formation and evolution of stars and galaxies, and the President Grant for Support of Leading
Scientific Schools NSh-5440.2012.2.

\end{document}